\documentclass[11pt,a4paper,english,showpacs,superscriptaddress,aps,prd]{revtex4-1}
\usepackage{amssymb}
\usepackage{amsmath}
\usepackage{graphicx}
\usepackage{babel}
\usepackage{hyperref}
\usepackage{color}

\begin{document}

\title{Supersymmetry breaking in the three-dimensional nonlinear sigma model}

\author{A.~C.~Lehum}
\email{andrelehum@ect.ufrn.br}
\affiliation{Instituto de F\'{\i}sica, Universidade de S\~{a}o Paulo\\
 Caixa Postal 66318, 05315-970, S\~{a}o Paulo, S\~{a}o Paulo, Brazil}
\affiliation{Escola de Ci\^encias e Tecnologia, Universidade Federal do Rio Grande do Norte\\
Caixa Postal 1524, 59072-970, Natal, Rio Grande do Norte, Brazil}

\author{A.~J.~da~Silva}
\email{ajsilva@fma.if.usp.br}
\affiliation{Instituto de F\'{\i}sica, Universidade de S\~{a}o Paulo\\
 Caixa Postal 66318, 05315-970, S\~{a}o Paulo, S\~{a}o Paulo, Brazil}


\begin{abstract}

In this work we discuss the phase structure of a deformed ${\cal N}=1$ supersymmetric nonlinear sigma model in a three-dimensional space-time. The deformation is introduced by a term that breaks supersymmetry explicitly, through imposing a slightly different constraint to the fundamental superfields of the model. Using the tadpole method, we compute the effective potential at leading order in $1/N$ expansion. From the gap equations, i.e., conditions that minimize the effective potential, we observe that this model presents two phases as the ordinary model, with two remarkable differences: $1)$ the fundamental fermionic field becomes massive in both phases of the model, which is closely related to the supersymmetry breaking term; $2)$ the $O(N)$ symmetric phase presents a meta-stable vacuum.  

\end{abstract}

\maketitle

\section{Introduction}

The Nonlinear Sigma model (NLSM) was first proposed to investigate the interaction between pions and nucleons~\cite{GellMann:1960np}. In lower dimensional systems, it is used to describe several aspects of condensed matter physics, for example, applications to ferromagnets~\cite{Polyakov:1975rr,Brezin:1975sq,Brezin:1976qa,Bardeen:1976zh}. In addition, this model provides a very good theoretical laboratory containing an interesting phase structure and at same time shares with the wealth of more realistic theories, being a simple example of an asymptotically free theory~\cite{Friedan:1980jf,Hikami:1980hi}. Recently, was conjectured that the $O(6)$ Sigma model emerges as a scaling function in AdS/CFT correspondence~\cite{Alday:2007mf,Bajnok:2008it}. 

The $O(N)$ NLSM can be defined through the action
\begin{eqnarray}\label{eq0}
S=\int\!{d^Dx}~\Big{\{}\frac{1}{2}\phi_a \Box\phi_a \Big{\}}~,
\end{eqnarray}

\noindent
where the fields $\phi_a$ are constrained to satisfy $\phi_a^2=\dfrac{N}{g}$, $D$ is the dimension of the space-time and the index $a$ assume the values $1,2,...,N$. 

It is useful rewrite the $O(N)$ NLSM action implementing the constraint over $\phi_a$ by the use of Lagrange multiplier,
\begin{eqnarray}\label{eq0a} 
S=\int\!{d^Dx}~\Big{\{}\frac{1}{2}\phi_a \Box\phi_a +\sigma\left(\phi_a^2-\frac{N}{g}\right)\Big{\}}~, 
\end{eqnarray}

\noindent
where the field $\sigma$ is the Lagrange multiplier that constraints $\phi_a^2=\dfrac{N}{g}$.

In the late of 1970's the phase structure and the renomalizability of the three-dimensional NLSM was established showing that this model possesses two phases~\cite{Arefeva:1979bd,Arefeva:1978fj}. One phase is $O(N)$ symmetric and exhibits a spontaneous generation of mass due to a non-vanishing vacuum expectation value (VEV) of the Lagrange multiplier field $\sigma$, i.e., $\langle\sigma\rangle\ne0$. On the other hand, if the fundamental bosonic field $\phi$ acquires a non-vanishing VEV, the $O(N)$ symmetry is spontaneously broken to $O(N-1)$, without any generation of mass. Several extensions of this model was after studied showing no changing in its phase structure~\cite{Arefeva:1980ms,Rosenstein:1989sg,Koures:1990hc,Koures:1991zu,Girotti:2001gs,Girotti:2001ku,Matsuda:1996vq,Jack:2001cd}. 

The 3D supersymmetric (SUSY) NLSM, in components~\cite{Koures:1990hc}, using the superfield formalism~\cite{Koures:1991zu}, and their noncommutative extensions~\cite{Girotti:2001gs,Girotti:2001ku}, was shown to be renormalizable to all orders in $1/N$ expansion. The phase structure of this model was also studied in \cite{Matsuda:1996vq}. In all these papers, a similar conclusion was achieved: no supersymmetry breaking is detected at leading order in $1/N$ expansion. 

The aim of this work is to show that imposing a more general constraint on the SUSY NLSM, the solutions that minimize the effective potential present broken supersymmetry at leading order in the $1/N$ expansion. Moreover, the $O(N)$ symmetric phase presents a meta-stable vacuum.

\section{Supersymmetric Nonlinear Sigma Model}

The usual three-dimensional ${\cal N}=1$ SUSY NLSM is defined through the action
\begin{eqnarray}\label{eq1}
S=\int\!{d^5z}~\Big{\{}\frac{1}{2}\Phi_a(z) D^2\Phi_a(z) +\Sigma(z)\left[\Phi_a(z)^2-\frac{N}{g}\right]\Big{\}}~,
\end{eqnarray}

\noindent
where $\Sigma$ is the Lagrange multiplier superfield that constraints $\Phi_a$ to satisfy $\Phi_a^2(z)=\dfrac{N}{g}$. With signature $(-,+,+)$, we are using notations and conventions as in~\cite{Gates:1983nr}. Such definitions and some useful identities can be found in the Supplemental Material \cite{supmat}.

The superfields appearing in this model possess the following $\theta$-expansion:
\begin{eqnarray}\label{eq1a}
&&\Phi_a(x,\theta)=\phi_a(x)+\theta^{\beta}\psi_{a\beta}(x)-\theta^2~F_a(x)~;\nonumber\\
&&\Sigma(x,\theta)=\rho(x)+\theta^{\beta}\chi_{\beta}(x)-\theta^2~\sigma(x)~.
\end{eqnarray}

We can see that the SUSY NLSM possesses more constraints than the non-supersymmetric one. Once the equation of motion of $\Sigma$ constraints
\begin{eqnarray}\label{constraints1}
\Phi_a^2(z)=\left[\phi_a^2+2\theta^{\beta}\phi_a\psi_{a\beta}-2\theta^2\left(\phi_aF_a-\dfrac{1}{2}\psi^{\beta}_a\psi_{a\beta}\right) \right]=\dfrac{N}{g}~,\nonumber
\end{eqnarray}

\noindent
it is easy to see that the component fields $\phi_a$, $\psi^{\alpha}_a$ and $F_a$ must satisfy
\begin{eqnarray}\label{constraints}
\phi_a^2 = \frac{N}{g}~,\hspace{1cm}\psi^\alpha_a\phi_a = 0~,\hspace{1cm} F_a\phi_a = \frac{1}{2}\psi^\beta_a\psi_{a\beta}~.
\end{eqnarray}

\noindent
Beyond the usual constraint $\phi_a^2=N/g$, the SUSY NLSM also exhibit the constraints $\psi^\alpha_a\phi_a=0$ and $F_a\phi_a=\dfrac{1}{2}\psi^\beta_a\psi_{a\beta}$.

Integrating the Eq.(\ref{eq1}) over $d^2\theta$, the action of the model can be cast as
\begin{eqnarray}\label{eq1b}
S&=&\int\!{d^3x}~\Big{\{}\frac{1}{2}\phi_a\Box \phi_a +\frac{1}{2}\psi^{\alpha}_{a}i{\partial_{\alpha}}^{\beta}\psi_{a\beta}
+\frac{1}{2}F_a^2+\sigma\left(\phi_a^2-\frac{N}{g}\right)\nonumber\\
&&+2\rho\left(F_a\phi_a+\frac{1}{2}\psi^{\beta}_{a}\psi_{a\beta}\right)+2\chi^{\beta}\psi_{a\beta}\phi_a\Big{\}}.
\end{eqnarray}

\noindent Notice that the usual model is obtained setting $\psi=\rho=\chi=0$, and the auxiliary field $\sigma$ must be non-vanishing.   

We can eliminate the auxiliary field $F_a$ using its equation of motion, $F_a=-2\rho\phi_a$. This way, the action
\begin{eqnarray}\label{eq1c}
S=\int\!{d^3x}\Big{\{}\frac{1}{2}\phi_a\Box \phi_a +\frac{1}{2}\psi^{\alpha}_{a}i{\partial_{\alpha}}^{\beta}\psi_{a\beta}
+\sigma\left(\phi_a^2-\frac{N}{g}\right)-2\rho^2\phi_a^2+\rho\psi_a^{\beta}\psi_{a\beta}
+2\chi^{\beta}\psi_{a\beta}\phi_a\Big{\}}~,
\end{eqnarray}

\noindent
describes the physical content of the model. It is easy to see that if exist a phase where mass is generated to the fundamental fields $\phi$ and $\psi$, their masses will be given by the VEV of the fields $\rho$ and $\sigma$ as
\begin{eqnarray}\label{eq1d}
M^2_{\phi}=4\langle\rho\rangle^2-2\langle\sigma\rangle~,\hspace{1cm} M^2_{\psi}=4\langle\rho\rangle^2~,
\end{eqnarray}
\noindent
from which we observe that SUSY should be spontaneously broken if $\langle\sigma\rangle\ne0$, as commented before.
For $\langle\sigma\rangle=0$ and for a non-vanishing VEV of $\rho$, the fundamental bosonic and fermionic fields acquire the same squared mass $4\langle\rho\rangle^2$, indicating generation of mass in a supersymmetric phase as is well-known~\cite{Koures:1990hc,Koures:1991zu,Girotti:2001gs,Girotti:2001ku,Matsuda:1996vq}. Here we find an intriguing point. While in the non-SUSY model the spontaneous generation of mass occurs due to $\sigma$ acquire a non-vanishing vacuum expectation value, in the SUSY version the field that acts like a "mass generator" to the fundamental fields is $\rho$, which is not present in the non-SUSY model. There is no soft transition or anything that we can interpret as a non-SUSY limit of the spontaneous generation of mass from the SUSY model.

Now, let us define a slightly deformed SUSY NLSM by
\begin{eqnarray}\label{eq1aa}
S=\int\!{d^5z}~\Big{\{}\frac{1}{2}\Phi_a(z) D^2\Phi_a(z) +\Sigma(z)\left[\Phi_a(z)^2-\frac{N}{g}\delta(z)\right]\Big{\}}~,
\end{eqnarray}

\noindent
with the single difference that $\Sigma$ is a Lagrange multiplier superfield that constraints $\Phi_a$ to satisfy $\Phi_a^2(z)=\dfrac{N}{g}\delta(z)$, where $\delta(z)$ is a constant superfield which possess the $\theta$-expansion $\delta(z)=\delta_1-\theta^2~g\delta_2$. Doing $\delta_2=0$ and $\delta_1=1$ we obtain the usual supersymmetric action for the SUSY NLSM Eq.(\ref{eq1}).

The equation of motion of the Lagrange multiplier superfield $\Sigma$ obtained from Eq.(\ref{eq1aa}) generates new constraints to the components of the fundamental superfields $\Phi_a$, namely
\begin{eqnarray}\label{constraints2}
\phi_a^2 = \frac{N}{g}\delta_1~,\hspace{1cm}\psi^\alpha_a\phi_a = 0~,\hspace{1cm} F_a\phi_a = \frac{1}{2}\psi^\beta_a\psi_{a\beta}+g\delta_2~.
\end{eqnarray}

To study the phase structure of the model, let us assume that the $\Sigma$ and the N-th component $\Phi_N(x,\theta)$ have a constant non-trivial VEV given by
\begin{eqnarray}\label{eq2}
\langle\Sigma\rangle&=&\Sigma_{cl}=\rho_{cl}-\theta^2\sigma_{cl}~,\nonumber\\
\langle\Phi_N\rangle&=&\sqrt{N}~\Phi_{cl}=\sqrt{N}~(\phi_{cl}-\theta^2F_{cl})~.
\end{eqnarray}

\noindent
Therefore, let us dislocate these superfields by $\Sigma\rightarrow(\Sigma+\Sigma_{cl})$ and $\Phi_N\rightarrow \sqrt{N}(\Phi_N+\Phi_{cl})$. So, we can rewrite the action Eq.(\ref{eq1}) in terms of the new fields as
\begin{eqnarray}\label{eq3}
S&=&\int\!{d^5z}\Big{\{}\frac{1}{2}\Phi_a (D^2+2\Sigma_{cl})\Phi_a 
+\Sigma\left(\Phi_a^2+N\Phi_{cl}^2+2{N}\Phi_{cl}\Phi_N-\frac{N}{g}\delta\right)\nonumber\\
&&+{N}\Phi_N\left(D^2\Phi_{cl}+2\Phi_{cl}\Sigma_{cl}\right)
+\frac{N}{2}\Phi_{cl} D^2\Phi_{cl}+N\Sigma_{cl}\left(\Phi_{cl}^2-\frac{1}{g}\right)\Big{\}}~.
\end{eqnarray}

We can note that the VEV of the superfield $\Sigma$, $\Sigma_{cl}$, give mass to the fundamental superfields $\Phi_a$. This ``mass'' is $\theta$-dependent, generating different masses to the bosonic and fermionic components of the superfield $\Phi_a$, showing a possible phase where supersymmetry is broken.

At leading order, the propagator of $\Phi_a$ superfield must satisfy the following equation
\begin{eqnarray}\label{eq4}
[D^2(z_1)+2\Sigma_{cl}]\Delta(z_1-z_2)=i\delta^{(5)}(z_1-z_2)~,
\end{eqnarray}

\noindent
where $\delta^{(5)}(z_1-z_2)\equiv \delta^{(3)}(x_1-x_2)\delta^{(2)}(\theta_1-\theta_2)$, and $\delta^{(2)}(\theta)=-\theta^2$. 

The solution to the above equation can be obtained from the ansatz
\begin{eqnarray}\label{eq5}
\Delta(z_1-z_2)&=& \left(C_1-\theta_1^2~C_2-\theta_2^2~C_3+\theta_1^{\alpha}\theta_2^{\beta}~\Delta_{\alpha\beta}
+\theta_1^2\theta_2^2~C_4\right)~\delta^{(3)}(x_1-x_2)~,
\end{eqnarray}

\noindent
where after some algebraic manipulations we can write the propagator of $\Phi_a$ superfield as
\begin{eqnarray}\label{eq8b}
\Delta(k)&=&-i\frac{D^2_{1}-2\rho_{cl}}{k^2+4\rho_{cl}^2}\Big{\{}
1+2\sigma_{cl}\frac{\delta^{(2)}(\theta_1)(D^2_{1}+2\rho_{cl})}{k^2+(4\rho_{cl}^2-2\sigma_{cl})}
\Big{\}}\delta^{(2)}(\theta_1-\theta_2)~.
\end{eqnarray}

\noindent
Notice that for $\sigma_{cl}=0$, the above propagator reduces to the usual propagator of a massive scalar superfield. A propagator presenting a similar form  was obtained in~\cite{Boldo:1999nd}. See Supplemental Material \cite{supmat} for details in obtaining the superfield propagator.

From Eq.(\ref{eq3}) we can see that exist a mixing between $\Phi_N$ and $\Sigma$, but this mixing only contributes to next-to-leading order in $1/N$ expansion. For now, we can neglect this mixing, since we will deal with the SUSY NLSM at leading order in $1/N$.

With the propagator of $\Phi_a$ superfield, let us evaluate the effective potential through the tadpole method~\cite{Weinberg:1973ua,Miller:1983fe,Miller:1983ri}. At leading order, the tadpole equation for $\Phi_N$ superfield can be cast as
\begin{eqnarray}\label{eq9}
N\left[D^2\phi_{cl}+2\Phi_{cl}\Sigma_{cl}\right]=N\left[F_{cl}+2\phi_{cl}\rho_{cl}-2\theta^2(\phi_{cl}\sigma_{cl}+F_{cl}\rho_{cl})\right].
\end{eqnarray}

\begin{figure}[b]
\includegraphics[height=4cm ,angle=0 ,width=8cm]{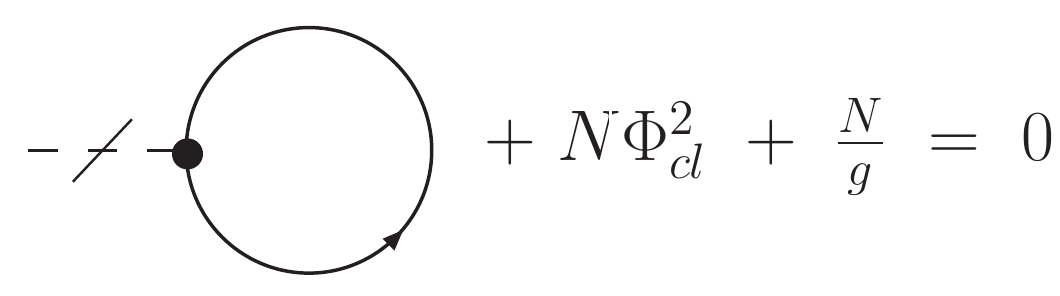}
\caption{Tadpole equation of $\Sigma$ at leading order. Continuous lines represent the $\Phi_a$ superfield propagator, while cut dashed line a removed external $\Sigma$ propagator.}\label{gap}
\end{figure}

On the other hand, the tadpole equation for $\Sigma$, Figure \ref{gap}, is $\displaystyle{\left[N\Phi_{cl}^2-\frac{N}{g}\delta+N\int\!\frac{d^3k}{(2\pi)^3}\Delta(k)\right]}$. Substituting the expression for $\Delta(k)$, and using the fact that $D^2\delta^{(2)}(\theta-\theta)=1$ and $\delta^{(2)}(\theta-\theta)=0$, we obtain
\begin{eqnarray}\label{eq11}&&N\Phi_{cl}^2-\frac{N}{g}\delta-iN\int\!\frac{d^3k}{(2\pi)^3}\Big{\{}
\frac{1}{k^2+(4\rho_{cl}^2-2\sigma_{cl})}+\frac{8\sigma_{cl}\rho_{cl}~\theta^2}{[k^2+(4\rho_{cl}^2-2\sigma_{cl})](k^2+4\rho_{cl}^2)}\Big{\}}\nonumber\\
&=&N\left[\phi_{cl}^2-\left(\frac{\delta_1}{g}-\frac{1}{g_c}\right)-\frac{\sqrt{4\rho^2_{cl}-2\sigma_{cl}}}{4\pi}-\theta^2\left(2\phi_{cl}F_{cl}-\frac{2}{\pi}\rho_{cl}^{3/2}+\frac{\rho_{cl}}{\pi}\sqrt{4\rho_{cl}^2-2\sigma_{cl}}+\delta_2\right)\right],
\end{eqnarray}

\noindent where $\dfrac{1}{g_c}$ is defined as usual $\displaystyle{\int_{\Lambda}\dfrac{d^3k}{(2\pi)^3}\dfrac{1}{k^2}}$. The coupling $g_c$ is the critical value of $g$ for that the NLSM exhibits the phase transition.

With the tadpole equations in the hand, the effective potential is obtained integrating Eq.(\ref{eq9}) over $\Phi_N$ and Eq.(\ref{eq11}) over $\Sigma$ as 
\begin{eqnarray}\label{eq12a}
\frac{V_{eff}}{N}&=&-\int{d^2\theta}\Big{\{}\int{d\Phi_N}\left[F_{cl}+2\phi_{cl}\rho_{cl}-2\theta^2(\phi_{cl}\sigma_{cl}+F_{cl}\rho_{cl})\right]\nonumber\\
&+&\int{d\Sigma}\left[\phi_{cl}^2-\lambda-\frac{\sqrt{4\rho^2_{cl}-2\sigma_{cl}}}{4\pi} -\theta^2\left(2\phi_{cl}F_{cl}-\frac{2}{\pi}\rho_{cl}|\rho_{cl}| +\frac{\rho_{cl}}{\pi}\sqrt{4\rho_{cl}^2-2\sigma_{cl}}+\delta_2\right)\right]\Big{\}}\nonumber\\
&=&-\frac{F_{cl}^2}{2}-\sigma_{cl}\left(2\phi_{cl}^2-\lambda\right)-6F_{cl}\rho_{cl}\phi_{cl} +\frac{2}{3\pi}(\rho_{cl}^2)^{3/2}-\frac{4}{3\pi}\left(\rho_{cl}^2-\frac{\sigma_{cl}}{2}\right)^{3/2}-\delta_2\rho_{cl}+C~,
\end{eqnarray}

\noindent
where $C$ is a constant of integration to be adjusted through the conditions that minimize the effective potential, the gap equations, and $\lambda\equiv\left(\dfrac{\delta_1}{g}-\dfrac{1}{g_c}\right)$ is a parameter that can be positive, negative or zero. In the thermodynamics of NLSM $\lambda$ is interpreted as a quantity proportional to magnetization of the system~\cite{Rosenstein:1989sg}. 

Looking to the tadpole equations in Eq.(\ref{eq9}) and Eq.(\ref{eq11}), we observe that the VEV's must to satisfy the following conditions:
\begin{eqnarray}\label{eq12}
&&F_{cl}+2\phi_{cl}\rho_{cl}=0~,\hspace{3cm} 
F_{cl}\rho_{cl}+\phi_{cl}\sigma_{cl}=0~,\nonumber\\
&&\phi_{cl}^2-\lambda-\frac{1}{2\pi}\sqrt{\rho_{cl}^2-\frac{\sigma_{cl}}{2}}=0~,\hspace{1cm}
\phi_{cl}F_{cl}+\frac{\rho_{cl}}{\pi}\left(\sqrt{\rho_{cl}^2-\frac{\sigma_{cl}}{2}}-|\rho_{cl}|\right)+\dfrac{\delta_2}{2}=0~.
\end{eqnarray}

Therefore, setting $\displaystyle{C=\left[\sigma_{cl}\phi^2_{cl}+4F_{cl}\rho_{cl}\phi_{cl} +\frac{2}{3\pi}\left(\rho_{cl}^2-\frac{\sigma_{cl}}{2}\right)^{3/2}\right]}$, the effective potential can be cast as
\begin{eqnarray}\label{eq12b}
\frac{V_{eff}}{N}&=&-\frac{F_{cl}^2}{2}-\sigma_{cl}\left(\phi_{cl}^2-\lambda\right)-2F_{cl}\rho_{cl}\phi_{cl} +\frac{2}{3\pi}(\rho_{cl}^2)^{3/2}-\frac{2}{3\pi}\left(\rho_{cl}^2-\frac{\sigma_{cl}}{2}\right)^{3/2}-\delta_2\rho_{cl}.
\end{eqnarray} 

As we did for the classical action, we can eliminate the auxiliary field $F_{cl}$ using its equation of motion, 
\begin{eqnarray}\label{eomF}
F_{cl}=-2\rho_{cl}\phi_{cl},
\end{eqnarray}

\noindent allowing us to write the effective potential as 
\begin{eqnarray}\label{eq12ccc}
\frac{V_{eff}}{N}&=&-\sigma_{cl}\left(\phi_{cl}^2-\lambda\right)+2\rho_{cl}^2\phi_{cl}^2 +\frac{2}{3\pi}\left[(\rho_{cl}^2)^{3/2}-\left(\rho_{cl}^2-\frac{\sigma_{cl}}{2}\right)^{3/2}\right]-\delta_2\rho_{cl}.
\end{eqnarray}

From the effective potential Eq.(\ref{eq12ccc}), the conditions that extremize the effective potential are given by
\begin{eqnarray}\label{gap-eq}
&&\phi_{cl}\left(\rho_{cl}^2- \frac{\sigma_{cl}}{2} \right)=0~,\nonumber\\
&&\phi_{cl}^2-\lambda-\frac{1}{2\pi}\sqrt{\rho_{cl}^2-\frac{\sigma_{cl}}{2}}=0~,\\
&&\rho_{cl}\left(2\pi\phi_{cl}^2+|\rho_{cl}|-  \sqrt{\rho_{cl}^2-\frac{\sigma_{cl}}{2}}\right)=\dfrac{\pi}{2}\delta_2.\nonumber
\end{eqnarray}

Solving these equations, we determine the field configurations that extremize the effective potential. Such solutions are presented in two phases, one $O(N)$ symmetric phase and another $O(N)$ broken to $O(N-1)$. The $O(N)$ symmetric phase,  $\lambda<0$ or $g>g_c$, the solutions are given by:
\begin{eqnarray} 
&&\phi_{cl}=0, \hspace{.3cm} \rho_{cl}=\pi|\lambda|+\frac{1}{2}\sqrt{2\pi(2\pi\lambda^2-\delta_2)}~,\hspace{.3cm}
\sigma_{cl}=\frac{1}{2}\left[2\pi|\lambda|+\sqrt{2\pi(2\pi\lambda^2-\delta_2)}\right]^2-8\pi^2\lambda^2~;\label{eq15b}\\
&&\phi_{cl}=0, \hspace{.3cm}  \rho_{cl}=-\pi|\lambda|-\frac{1}{2}\sqrt{2\pi(2\pi\lambda^2+\delta_2)}~,\hspace{.3cm}
\sigma_{cl}=\frac{1}{2}\left[2\pi|\lambda|+\sqrt{2\pi(2\pi\lambda^2+\delta_2)}\right]^2-8\pi^2\lambda^2.\label{eq15c}
\end{eqnarray}

\noindent Note for \emph{real} solutions, the parameter $\delta_2$ is constrained to be $|\delta_2|\le2\pi\lambda^2$. Moreover, as we will see, exist a $\delta_2\neq 0$ which $V_{eff}$ assumes its minimum value. Setting $\delta_2=0$ we have the well-known solutions~\cite{Koures:1990hc,Koures:1991zu,Girotti:2001gs,Girotti:2001ku,Matsuda:1996vq}
\begin{eqnarray}  \rho_{cl}=\pm 2\pi|\lambda|~,\hspace{1cm}\phi_{cl}=F_{cl}=\sigma_{cl}=0~.\label{eq15} \end{eqnarray}

\begin{figure}[b]
\includegraphics[height=4cm ,angle=0 ,width=8cm]{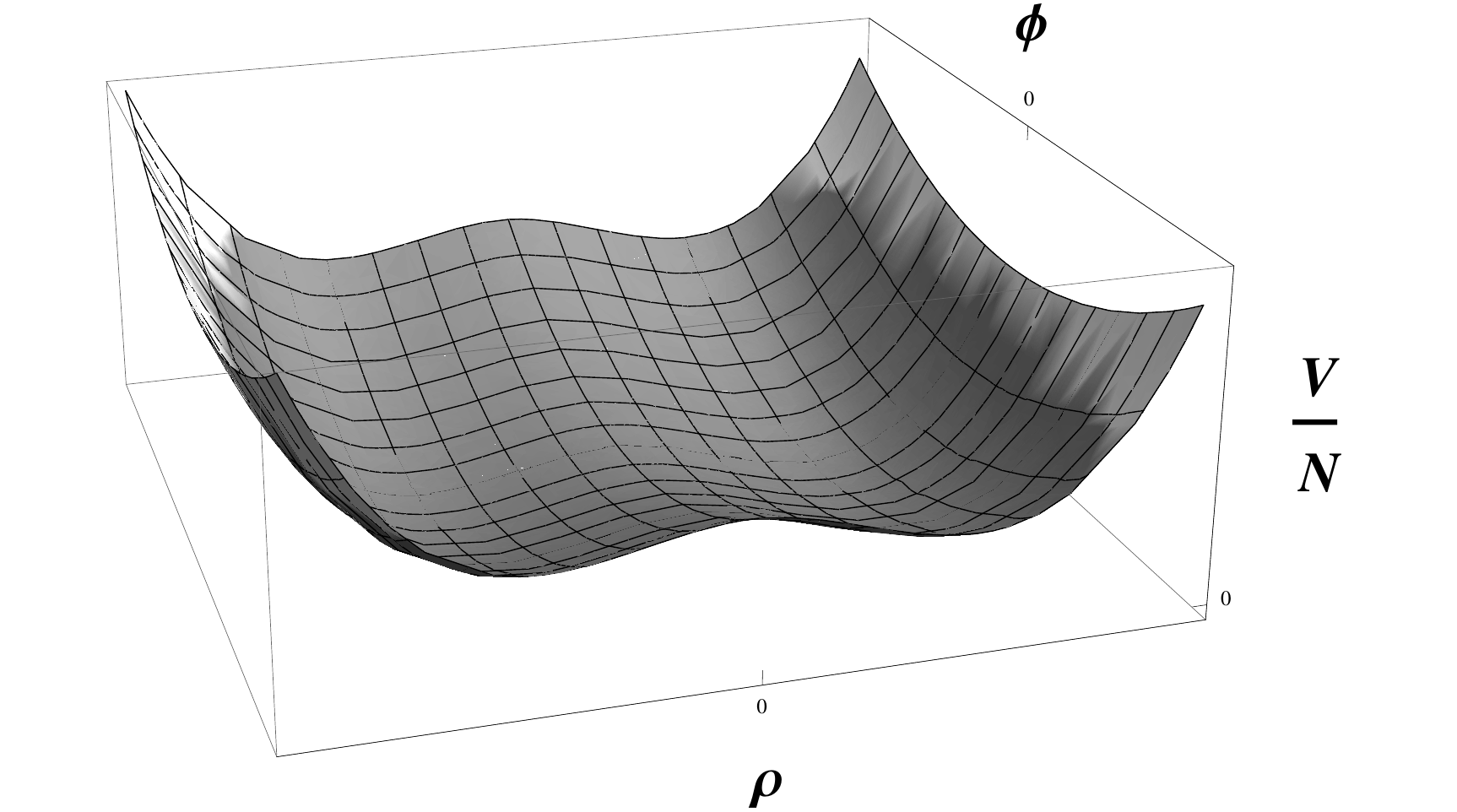}
\includegraphics[height=4cm ,angle=0 ,width=8cm]{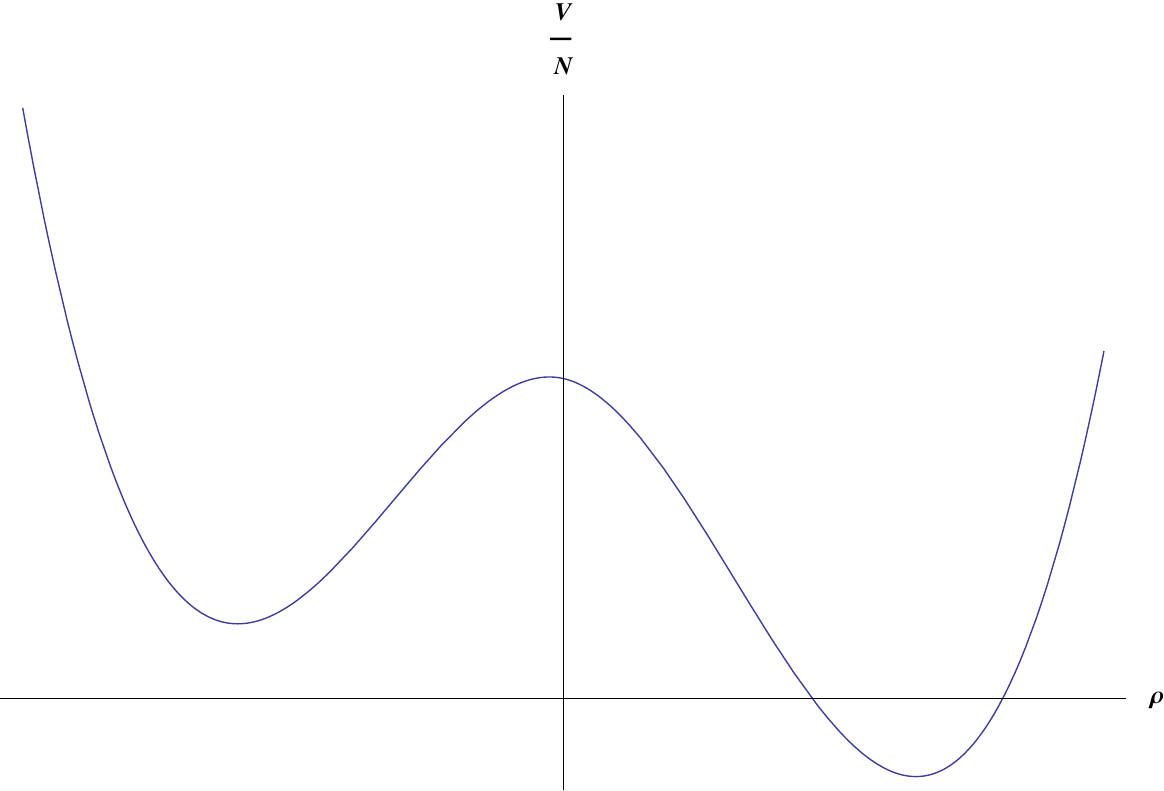}
\caption{Effective potential in the $O(N)$ symmetric phase as function of $\rho_{cl}$ and $\phi_{cl}$. The plot in the right side of the figure is a slice of the $V_{eff}$ at $\phi_{cl}=0$, evidencing the presence of a meta-stable vacuum.}\label{ons1}
\end{figure}

The solution Eq.(\ref{eq15b}) is the global minimum of the effective potential while Eq.(\ref{eq15c}) is a local one. The effective potential is plotted in the Figure \ref{ons1} as a function of $\rho_{cl}$ and $\phi_{cl}$, where it is possible to see the true and the false vacua. 

In the minimum, $V_{eff}$ is negative, this is because we are dealing with an explicit breaking of supersymmetry. The generated masses for the fundamental fields $\phi$ and $\psi$ in the $O(N)$ symmetric phase are given by   
\begin{eqnarray}\label{mass1}
M^2_{\phi}&=&4\langle\rho\rangle^2-2\langle\sigma\rangle=16\pi^2\lambda^2 \\
M^2_{\psi}&=&4\langle\rho\rangle^2=8\pi^2\lambda^2+4\pi|\lambda|\sqrt{2\pi(2\pi\lambda^2-\delta_2)}-2\pi\delta_2.
\end{eqnarray}

\noindent In  the limit $\delta_2\rightarrow0$ the masses $M^2_{\phi}=M^2_{\psi}$ and supersymmetry is restored.

The second phase, $O(N)$ symmetry is broken to $O(N-1)$, $\lambda>0$ or $g<g_c$, and the solutions that minimize the effective potential are given by
\begin{eqnarray}
&&\phi_{cl}=\pm\sqrt{\lambda} ~,\hspace{.5cm}   \rho_{cl}=\pi\lambda-\frac{1}{2}\sqrt{2\pi(2\pi\lambda^2-\delta_2)}~,\hspace{.5cm}
\sigma_{cl}=\frac{1}{2}\left[2\pi\lambda-\sqrt{2\pi(2\pi\lambda^2-\delta_2)}\right]^2~;\label{eq13a}\\
&&\phi_{cl}=\pm\sqrt{\lambda} ~,\hspace{.5cm} \rho_{cl}=-\pi\lambda+\frac{1}{2}\sqrt{2\pi(2\pi\lambda^2+\delta_2)}~,\hspace{.5cm}
\sigma_{cl}=\frac{1}{2}\left[2\pi\lambda-\sqrt{2\pi(2\pi\lambda^2+\delta_2)}\right]^2~.\label{eq13d}
\end{eqnarray}

\noindent where, just as $O(N)$ symmetric phase discussed before, for $\delta_2\rightarrow0$ the above solutions collapse to 
\begin{eqnarray}
&&\phi_{cl}=\pm\sqrt{\lambda}~,\hspace{1cm}F_{cl}=\sigma_{cl}=\rho_{cl}=0~.\label{eq13c}
\end{eqnarray}

Just as the supersymmetric and non-supersymmetric cases, in the $O(N)$ symmetric phase the scalar field $\phi$ is kept massless, i.e., $M^2_{\phi}=0$. But, due to the parameter that breaks supersymmetry, $\delta_2$, the fundamental fermion of the model acquires the mass  
\begin{eqnarray}\label{mass2}
M^2_{\psi}&=&4\langle\rho\rangle^2=\left[2\pi\lambda-\sqrt{2\pi(2\pi\lambda^2-\delta_2)}\right]^2.
\end{eqnarray}

\noindent It is easy to see that if $\delta_2\rightarrow0$ so $M^2_{\psi}\rightarrow0$. 

Finally, let us deal with the optimal value of the SUSY-breaking parameter $\delta_2$. Eliminating, from Eq.(\ref{eq12ccc}), all fields by the use of their equations of motion, except the fundamental field $\phi$, to $\lambda>0$ we find
\begin{eqnarray}\label{delta1}
\frac{V_{eff}}{N}&=&\frac{1}{6}\Big{\{}-12\pi\lambda(\delta_2+2\pi\lambda^2)-3(4\pi\lambda^2-\delta_2)\sqrt{2\pi(2\pi\lambda^2-\delta_2)}\nonumber\\
&&+\left[32\pi^4\lambda^2-8\pi^3\delta_2-16\pi^3\lambda\sqrt{2\pi(2\pi\lambda^2-\delta_2)}  \right]^{3/2}\nonumber\\
&&+144\pi^2\lambda\phi_{cl}^2(\lambda-\phi_{cl}^2)+48\pi^2\phi_{cl}^6-32\pi^2|\lambda-\phi_{cl}^2|\Big{\}}.
\end{eqnarray} 

\noindent
Minimizing  Eq.(\ref{delta1}) for $\delta_2$ we obtain the solution
\begin{eqnarray}\label{delta2}
\delta_2&=&\frac{3\pi}{2}\lambda^2.
\end{eqnarray} 

The effective potential Eq.(\ref{eq12ccc}) evaluated for $\delta_2=\frac{3\pi}{2}\lambda^2$ is given by
\begin{eqnarray}\label{delta3}
\frac{V_{eff}}{N}&=&-\sigma_{cl}\left(\phi_{cl}^2-\lambda\right)+2\rho_{cl}^2\phi_{cl}^2 +\frac{2}{3\pi}\left[(\rho_{cl}^2)^{3/2}-\left(\rho_{cl}^2-\frac{\sigma_{cl}}{2}\right)^{3/2}\right]-\frac{3\pi}{2}\lambda^2\rho_{cl}.
\end{eqnarray}

One interesting note is that $\delta_2=0$ becomes a local maximum in this model. Once introduced the SUSY-breaking parameter, the supersymmetric solutions are not the solutions that minimize the effective potential anymore.

\section{Final remarks}

Summarizing, the three-dimensional supersymmetric nonlinear sigma model, deformed by a non-supersymmetric constraint, possess two phases. In the first one is the $O(N)$ symmetric phase,  $\lambda<0$ or $g>g_c$, which possess the remarkable characteristic of the presence of a meta-stable vacuum. In this phase, all fields acquire a non-vanishing vacuum expectation value, generating masses to the fundamental fields $\phi$ and $\psi$. These masses are different for non-vanishing $\delta_2$, coupling responsible for supersymmetry breaking. In the limit $\delta_2\rightarrow0$ the masses of $\phi$ and $\psi$ tend to be equal, restoring the supersymmetry. In the $O(N)$ broken phase, only the components of the Lagrange multiplier superfield acquire a non-vanishing vacuum expectation value, generating mass to the fermionic field $\psi$ and keeping $\phi$ massless. Also in this phase, the limit $\delta_2\rightarrow0$ can be taken to restore the supersymmetric solutions. An important note is the fact that $\delta_2$ can not be chosen arbitrarily. It possesses an optimal value that minimizes the effective potential.       

Finally, we think that gauge and noncommutative extensions (with constant noncommutative parameter; see, for example the SUSY CP$^{(N-1)}$ model presented in Ref.~\cite{Ferrari:2006xx}) of this model should present similar structure, including the presence of the meta-stable vacuum, since in general the tadpole diagrams in noncommutative models are the same of the commutative ones.

\vspace{.5cm}
{\bf Acknowledgments.} This work was partially supported by the Brazilian agencies Funda\c{c}\~{a}o de Amparo \`{a} Pesquisa do Estado de S\~{a}o Paulo (FAPESP), Conselho Nacional de Desenvolvimento Cient\'{\i}fico e Tecnol\'{o}gico (CNPq) and Funda\c{c}\~{a}o de Apoio \`{a} Pesquisa do Rio Grande do Norte (FAPERN).

\end{document}